\documentclass{aa}  
\usepackage{graphicx}
\usepackage{txfonts}
\begin{document}

   \title{Connection between the Long Secondary Period\\ phenomenon and the red giant evolution}
   \titlerunning{Connection between the LSP phenomenon and the red giant evolution}

   \author{Micha{\l} Pawlak}
          
    \authorrunning{M. Pawlak}

   \institute{Institute of Theoretical Physics, Faculty of Mathematics and Physics, Charles University in Prague, Czech Republic\\
   Astronomical Observatory, Jagiellonian University, ul. Orla 171, 30-244 Krak{\'o}w, Poland\\
   Astronomical Observatory, University of Warsaw, Al. Ujazdowskie 4, 00-478, Warszawa, Poland\\
              \email{michal.pawlak@utf.mff.cuni.cz}
             }

   \date{}

 
  \abstract
   {}
   {The mechanism behind the Long Secondary Period (LSP) observed in pulsating red giants still remains unknown. In this work, I investigate the connection between the Red Giant Branch and Asymptotic Giant Branch evolution and the appearance of the LSP - the phenomenon observed in a large fraction the red giants. }
   {I use the OGLE-III sample of the OSARG variables in the Large Magellanic Cloud. I construct the density maps in the period-luminosity as well as color-magnitude planes for the stars showing LSP and compare them to the remaining giants. I also fit the spectral energy distribution to test whether an additional source of reddening is present in the LSP stars.}
   {I post a hypothesis that the LSP phenomenon may be related to a transition between the different pulsation period-luminosity sequences. I also show that an overabundance of the stars showing Long Period Variables can be observed around the Tip of the Red Giant Branch, and much more prominently, at the upper part of the Asymptotic Giant Branch. The main over-density region appears to be slightly fainter and redder than the bulk of the Asymptotic Giant Branch. It also seems to correspond to the area of the Hertzsprung–Russell diagram where stable winds and high mass loss are present.}
   {The LSP can possibly be a recurring phenomenon appearing and disappearing in various points of the red giant evolution. The LSP stars appear to be more reddened than other giants, which suggests the intrinsic nature of the reddening, likely related to large dust emission. The analysis seems to confirms the hypothesis that there is a relation between the mass loss due to the presence of strong stellar wind and the appearance of LSP. }


   \maketitle
%

\section{Introduction}

The Long Secondary Period (LSP) is a phenomenon that is observed for a large fraction of pulsating red giants, often referred to as Long Period Variables (LPVs). Even though LSP has been known for decades \citep{oconnell1933, paynegaposchkin1954, houk1963}, the mechanism behind it still remains not fully explained. LSP giants form the sequence D in the period-luminosity (PL) relation plane, which was identified for the first time by \citet{wood1999} and further investigated by \citet{derekas2006} and \citet{soszynskietal2007}. While the other PL sequences formed by LPVs can be explained either by radial pulsations (sequences A to C) or binarity of the close, ellipsoidal red giants (sequence E), the origin of the sequence~D remains unknown. LSP has been also observed in pulsating red supergiants \citep{yang2012}. 

A number of hypotheses to explain the LSP phenomenon have been proposed, including: pulsation, rotation, mass ejection, convective cells and binarity. First possibility is that LSP variability originates from pulsation. \citet{wood2000b} argues that, since no radial pulsation mode can have a period longer than the fundamental mode, due to the adiabatic theory therefore, if LSP is caused by pulsation, it must be non-radial. At the same time, \citet{wood2000b} points out that the typical non-radial pulsation modes have amplitudes which are too small to explain the amplitudes observed for the LSP stars. On the other hand, a peculiar model of pulsations connected to convection proposed by \citet{wood2000a} might by able to produce amplitudes that are high enough. The idea of non-radial pulsations as an explanation of LSP has been given as the most likely by \citet{hinkle2002} and \citet{wood2004}. Convective modes of non-radial pulsation were also investigated by \citet{saio2015}.

Another possible explanation, mentioned by \citet{wood2004}, is that LSP is caused by the rotation of a spotted star. \citet{olivier2003} investigated this theory, however concluded that the periods of LSP are too long to be related to stellar rotation, based on the analysis of the rotational velocities of a sample of LSP stars. \citet{takayama2015} explored the mass ejection and dust absorption as well as large spots on a rotating star as potential answers to the LSP puzzle, but concluded that neither of the mechanism is likely to be the correct solution.

Finally, LSP can be related to the binarity of a pulsating red giant, which is most likely accompanied by a much smaller and fainter companion. \citet{wood1999} argued that the binary scenario is the most likely explanation of LSP, since the variability amplitudes are to large for non-radial pulsations or rotational variability. \citet{soszynski2007} as well as \citet{soszynski2014} claim that the shape of the light curves of the LSP variables, support the binary hypothesis. It is also known that the sequence D, formed by the LSP stars, overlap with the sequence E formed by ellipsoidal binaries, and can be interpreted as its extension \citep{soszynski2004,derekas2006,soszynskietal2007}.

A spectroscopic analysis of a large sample of LSP stars done by \citet{nicholls2009} reveals that the radial velocities of the LSP stars are too low for a binary system with a secondary component of a mass similar to the primary. However, scenario with a much less massive, likely sub-stellar companion is still a possible solution. Non-radial pulsation theory also cannot be ruled out. 

\begin{figure*}
 \centering
	\includegraphics[bb=40 50 560 760, width=0.5\textwidth, angle=270, clip]{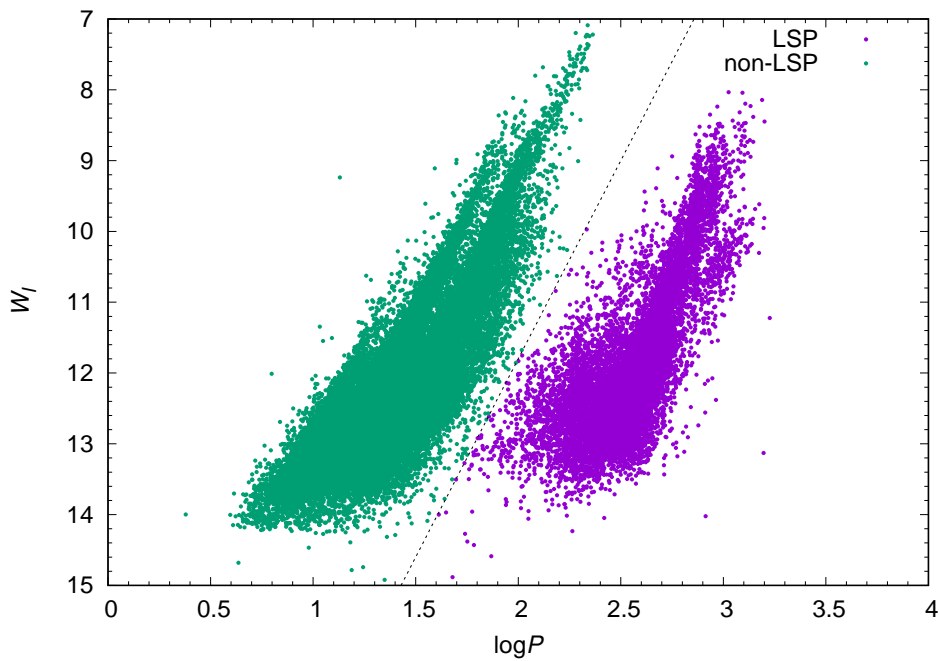} 
	\caption{Separation of the LSP OSARGs from the rest of the sample.}
	    \label{fig:sep}
\end{figure*}

Recent studies show that the onset of LSP is likely to be related to the evolution of the star on the Asymptotic Giant Branch (AGB). The pulsational period of the LPV star is getting longer as it evolves, and the dominant mode of pulsation shifts from the third overtone to the fundamental mode. This corresponds to moving to a subsequent pulsational sequence on the PL diagram \citep{wood2015, trabucchi2019}.  \citet{trabucchi2017} show that the appearance of the LSP coincides with the transition between sequences B and C' which are both first-overtone sequences. \citet{mcdonald2019} claims that a strong dust-driven wind, causing a mass loss of $\sim 10^7\times M_{\sun}$ is also tied to the same point at the AGB evolution.

In this paper I investigate the connection between the different stages of the Red Giant Branch RGB and AGB evolution.

\section{Analysis}

The analysis is based on the sample of LPV stars in the Large Magellanic Cloud from the OGLE-III catalog of variable stars \citep{soszynskietal2007}. For the technical details of the OGLE-III survey refer to \citet{udalski2003}. From the whole catalog of the LPVs, I take only the stars classified as OASRGs, in order to assure that the list of the LSPs in the analyzed sample is complete. For the OSARGs this can be easily achieved since the pulsational variability of these stars have typically lower amplitudes than the LSP variability, which means that the LSP will be detected as a dominant source of variability if it is present in a given star. The same statement is not true for the SRV stars which often show pulsation with amplitudes comparable or higher to LSP. The sample contains $78\:919$ OSARG variables. 

As a first step, I divide the sample of OGLE-III OSARGs into LSP and non-LSP. For that purpose I use the ${\log}P$-$W_I$ diagram, shown in Fig.~\ref{fig:sep}, where $P$ is the dominate period (the one with the highest signal-to-noise ratio) and $W_I$ is the Wesenheit index defined as in \citet{soszynskietal2007} as $W_I = I - 1.55 (V-I)$. Since the amplitude of the LSP variability is generally much higher than that of the pulsations, the LSP is usually detected as the dominant period for the stars that show this phenomenon. Therefore, the LSP OSARGs lie on the $D$ sequence on the right end of the ${\log}P$-$W_I$ diagram, while non-LSP OSARGs occupy the sequences $a_1$-$a_4$ and $b_1$-$b_2$ in the shorter periods area. The labels of the sequences are adopted from \citet{soszynskietal2007}. The stars with $W_I < -5.6{\log}P+23$ are flagged as LSP and the remaining ones as non-LSP. There might be a small contamination to the LSP sample by ellipsoidal variables from sequence E, which partially overlap with the sequence D.  For the LSP stars, I take the next strongest period from \citet{soszynskietal2007} as a pulsational period.
 
\begin{figure*}
 \centering
	\includegraphics[bb=100 50 480 800, width=0.405\textwidth, angle=270, clip]{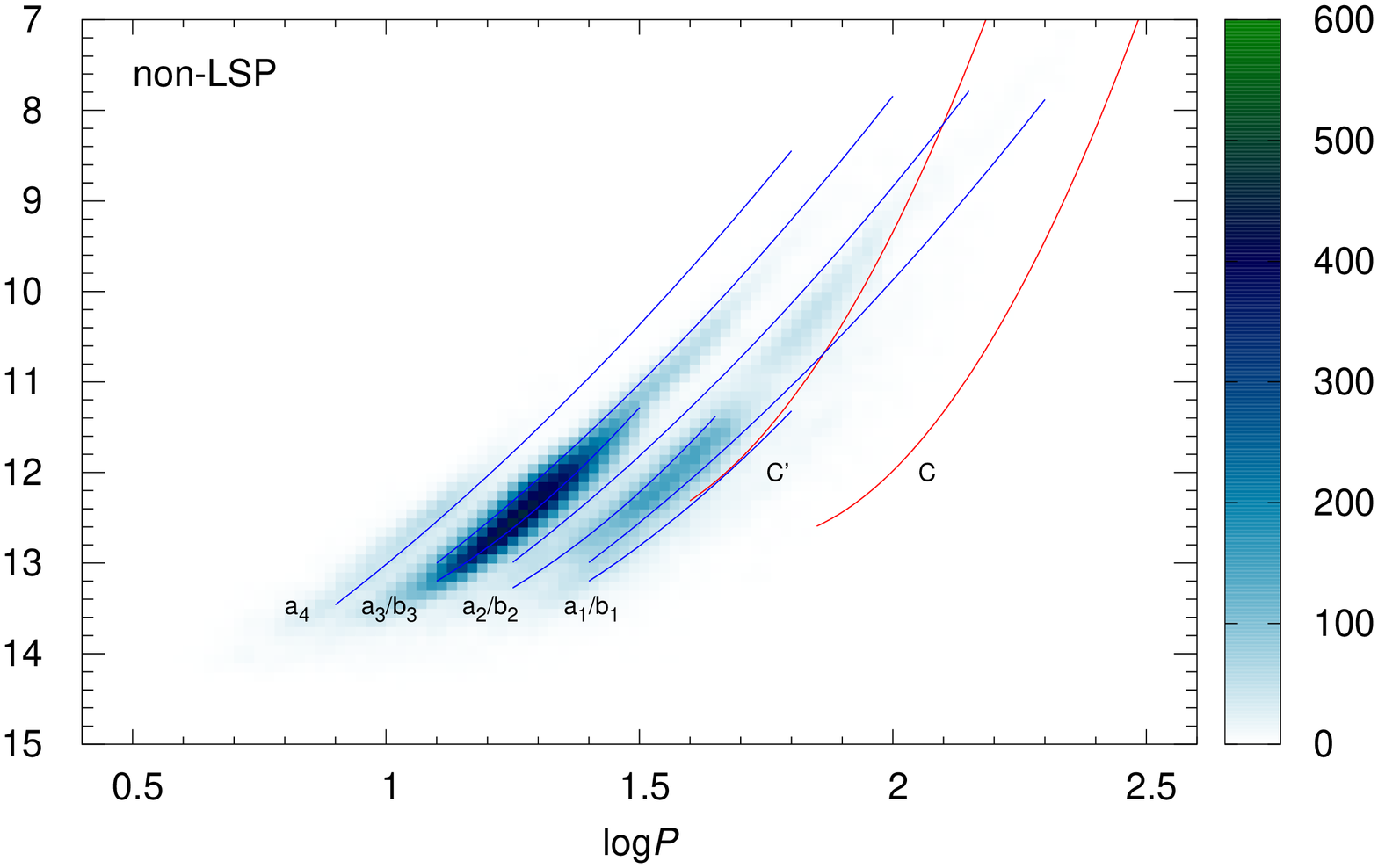} \\
	\includegraphics[bb=100 50 480 800, width=0.405\textwidth, angle=270, clip]{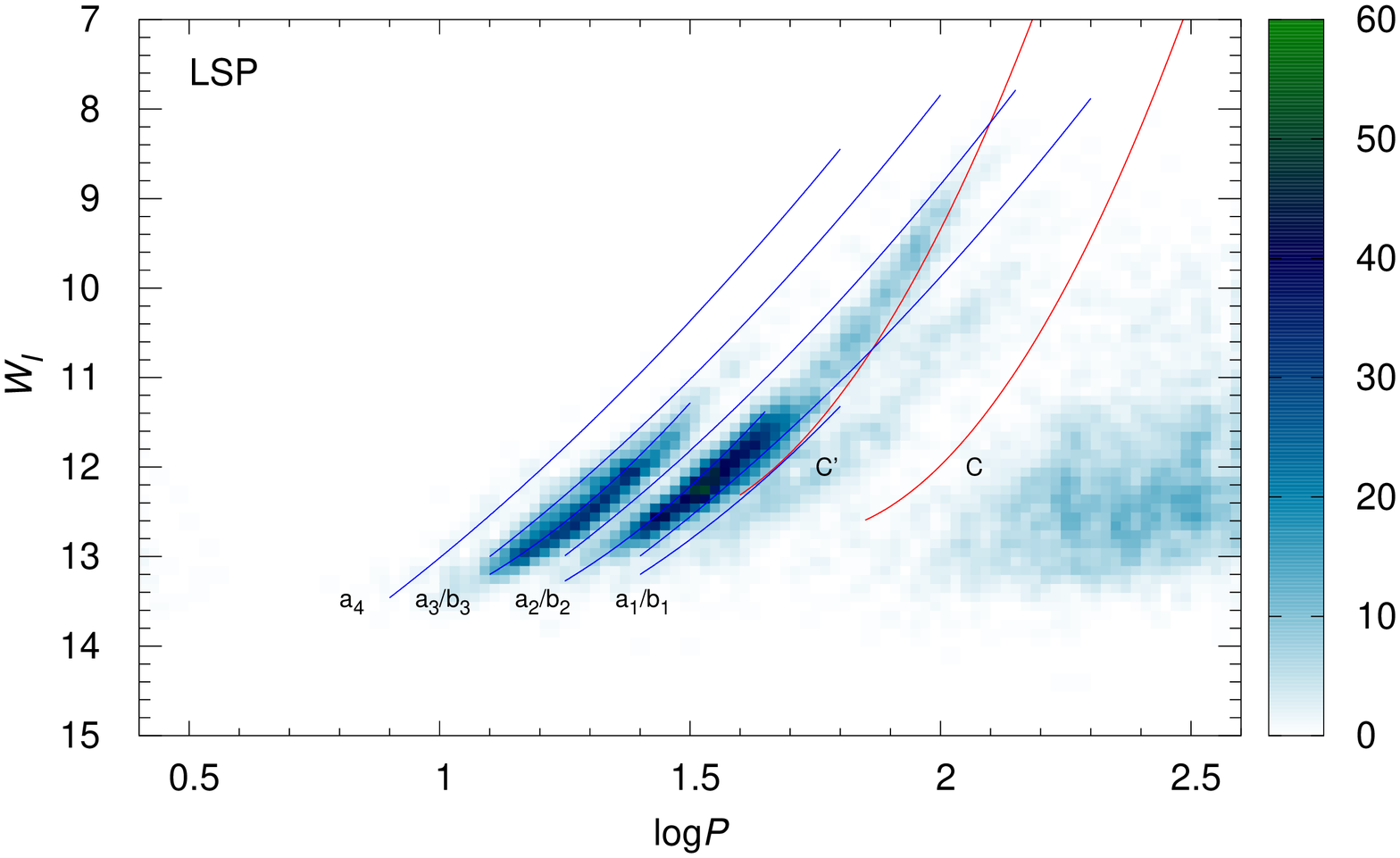} \\
	\includegraphics[bb=100 50 535 800, width=0.465\textwidth, angle=270, clip]{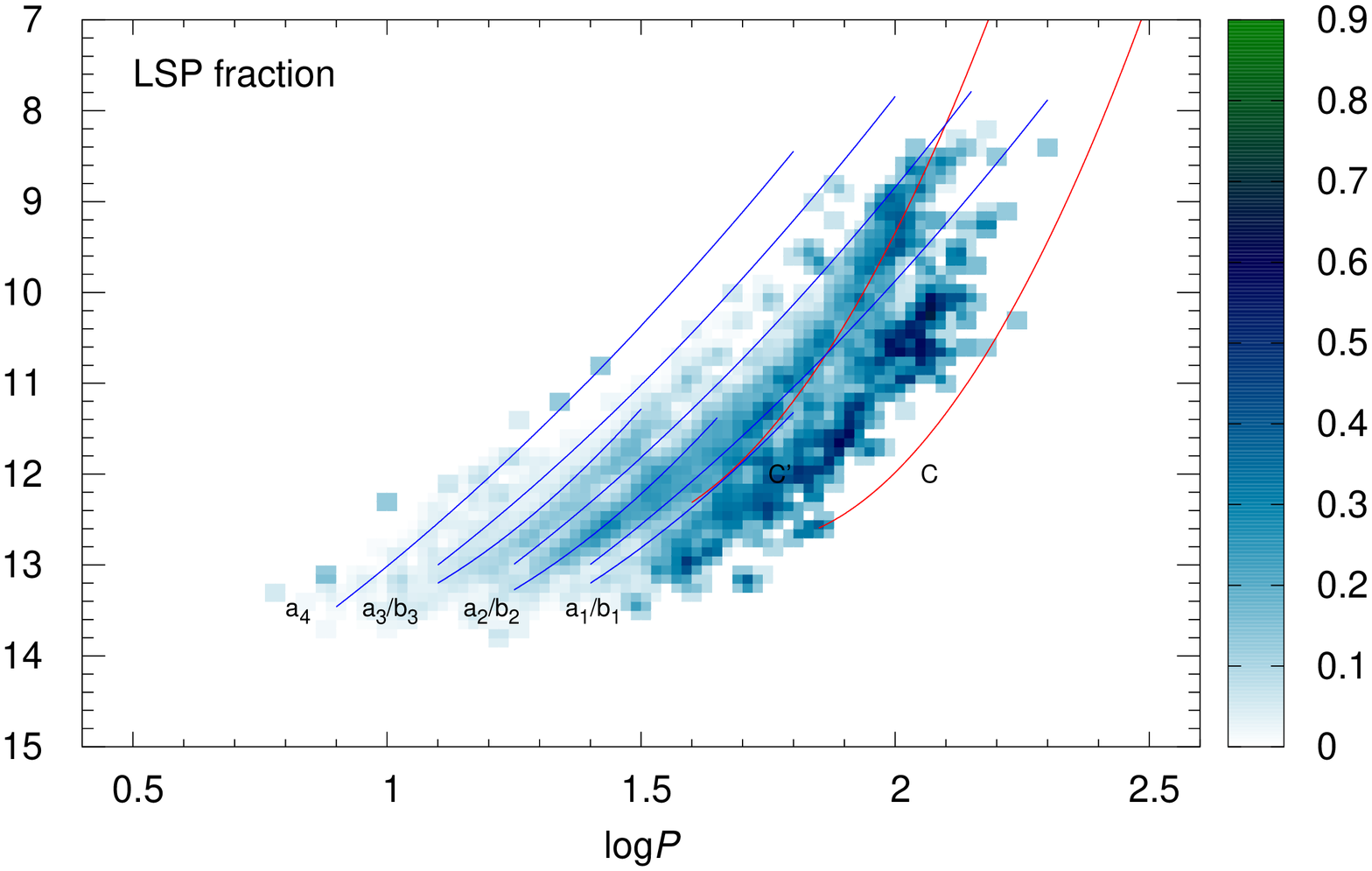}
   \caption{Density map in the PL plane for non-LSP OSARGs - upper panel, LSP OSARGs - middle panel and the fraction of OSARGs showing LSP - lower panel. In the middle and lower panels, the period used for the LSP OSARGs is the pulsational (i.e. second most prominent) period, not the LSP itself. PL relation for OSARGs and SRVs \citep{soszynski2007} are marked with blue and red lines respectively.}
    \label{fig:pl}
\end{figure*}

\begin{figure*}
 \centering
	\includegraphics[bb=100 50 480 800, width=0.405\textwidth, angle=270, clip]{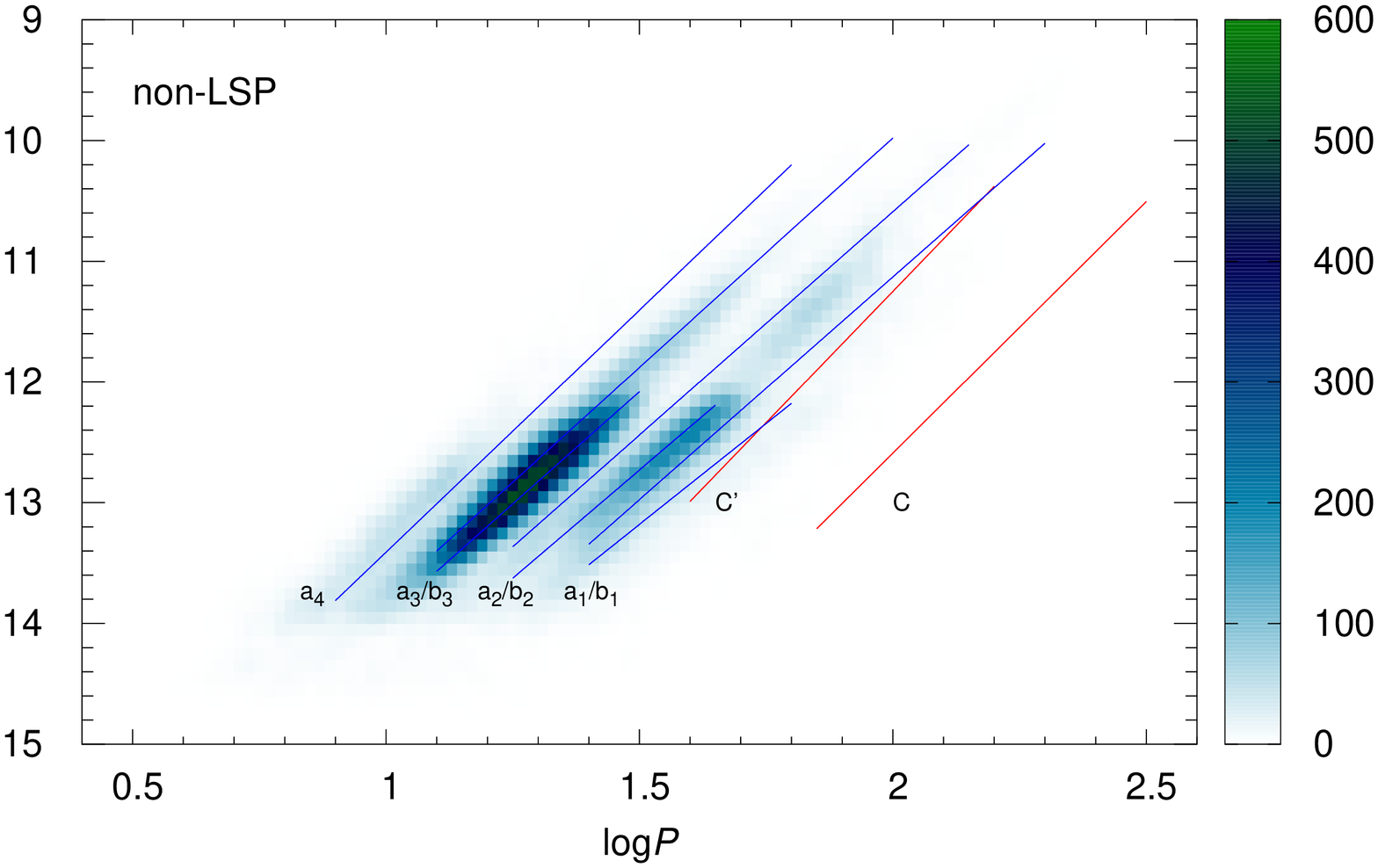} \\
	\includegraphics[bb=100 50 480 800, width=0.405\textwidth, angle=270, clip]{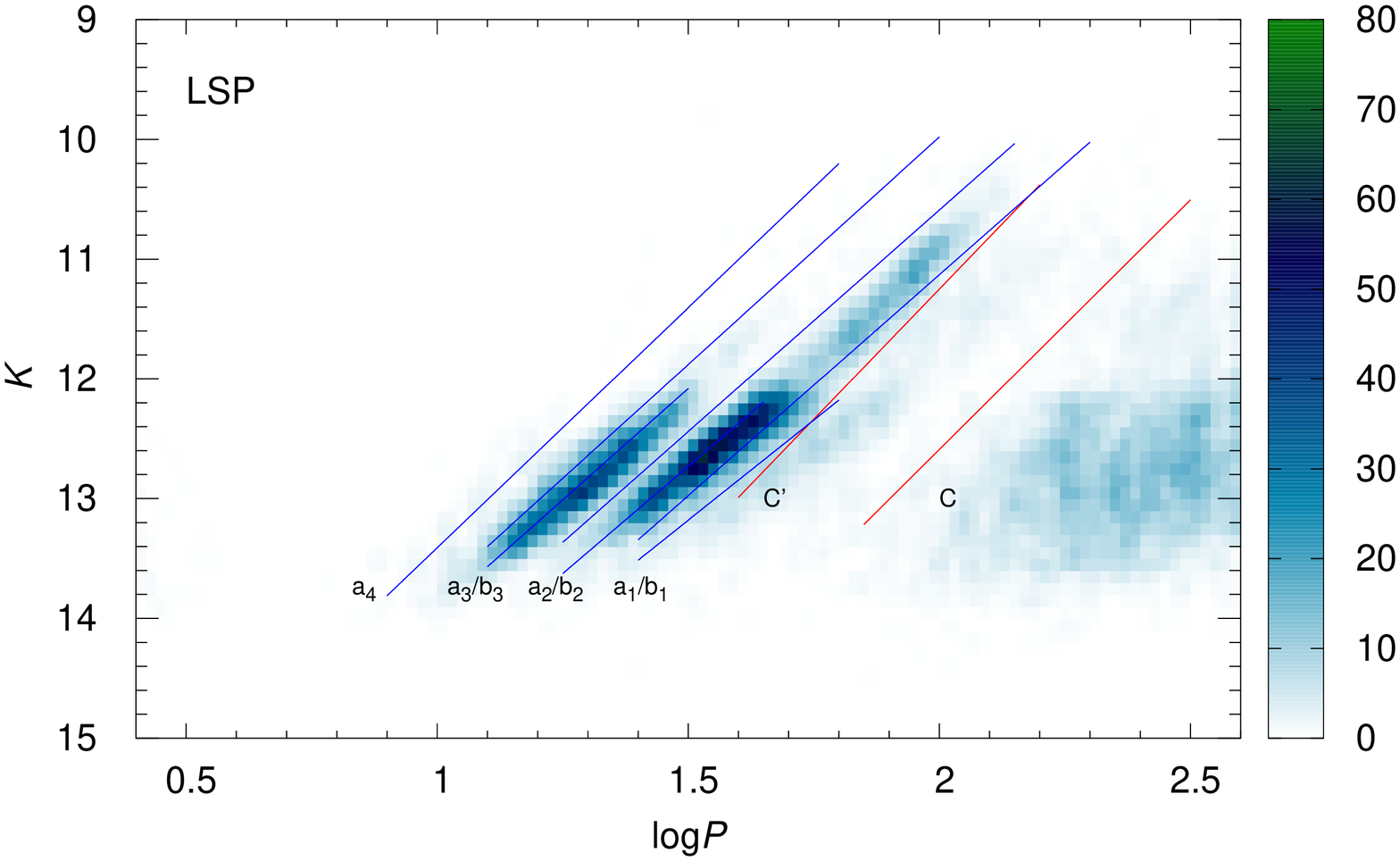} \\
	\includegraphics[bb=100 50 535 800, width=0.465\textwidth, angle=270, clip]{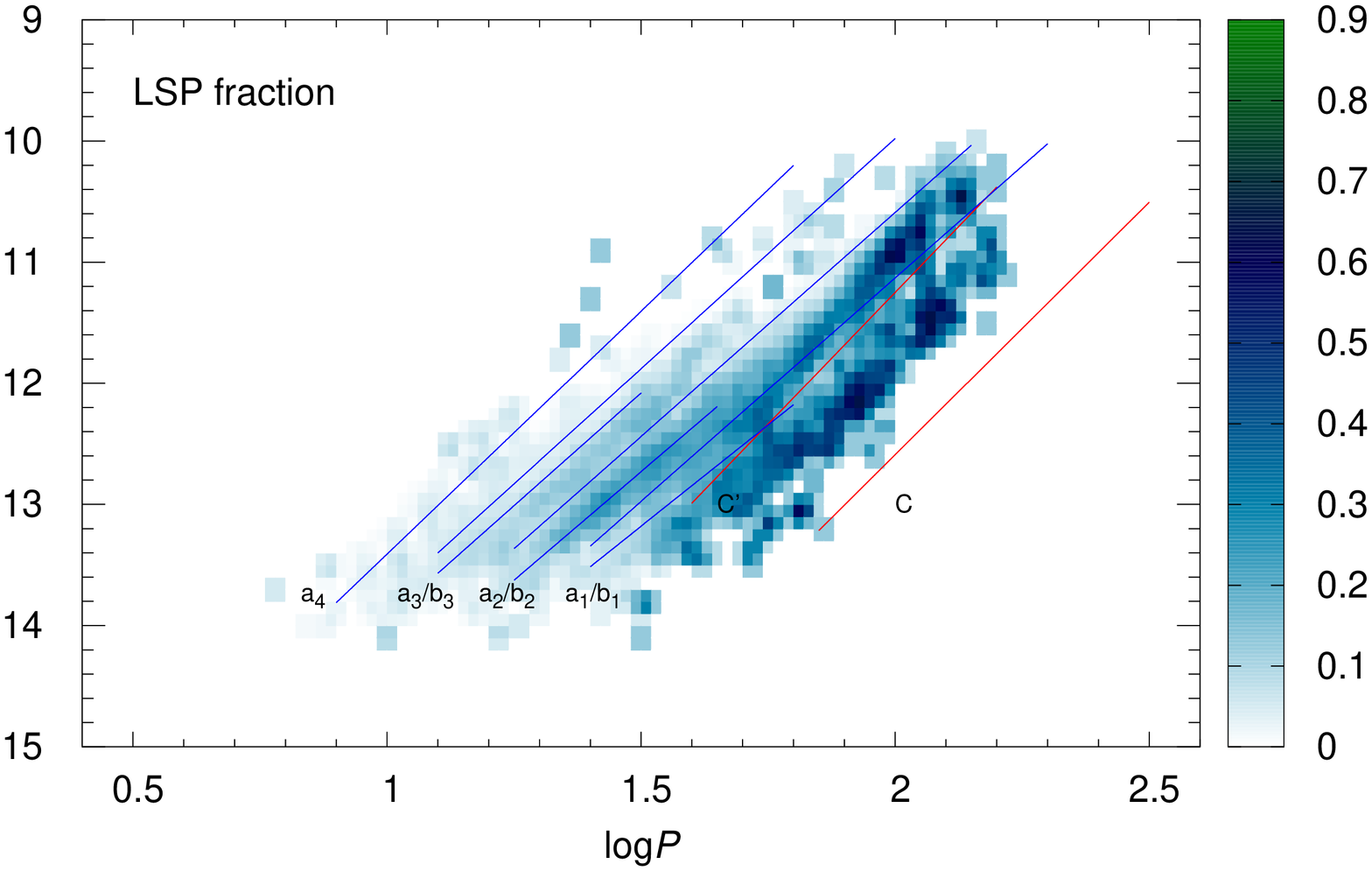}
   \caption{The same as Fig.~\ref{fig:pl} but in $K$-band.}
    \label{fig:plk}
\end{figure*}

 Fig.~\ref{fig:pl}~and~\ref{fig:plk} present the density map in the $\log P$ - $W_I$ and $\log P$ - $K$ planes, respectively, for the both non-LSP and LSP OSARGs. The map shows the raw number of stars per bin without any smoothing. The LSP stars are plotted with the short, pulsational period. The LPV period-luminosity sequences adopted from \citet{soszynskietal2007} are over-plotted with the solid lines. 

As expected, the vast majority of non-LSP stars (Fig.~\ref{fig:pl}~and~\ref{fig:plk} upper panel) clump at the sequences $b_1$ - $b_4$, especially the $b_3$ sequence. However, the picture for the LSP stars (Fig.~\ref{fig:pl}~and~\ref{fig:plk} middle panel) appears to be different. The majority of the LSPs lie in between the sequences, especially between $b_3$ and $b_2$ as well as between $b_2$ and $b_1$. There is also a broad region of LSP stars to the right from the sequence C. These are the stars for which the second strongest period is an alias of the LSP, not a pulsational period.  In the bottom panels of the Fig.~\ref{fig:pl}~and~\ref{fig:plk}, I show the relative density of the LSP stars as a fraction of all OSARGs. A clear over-density of LSPs between $b_1$/$C`$ and $C$ sequences can be observed here. 

\begin{figure*}
 \centering
	\includegraphics[bb=100 50 480 800, width=0.41\textwidth, angle=270, clip]{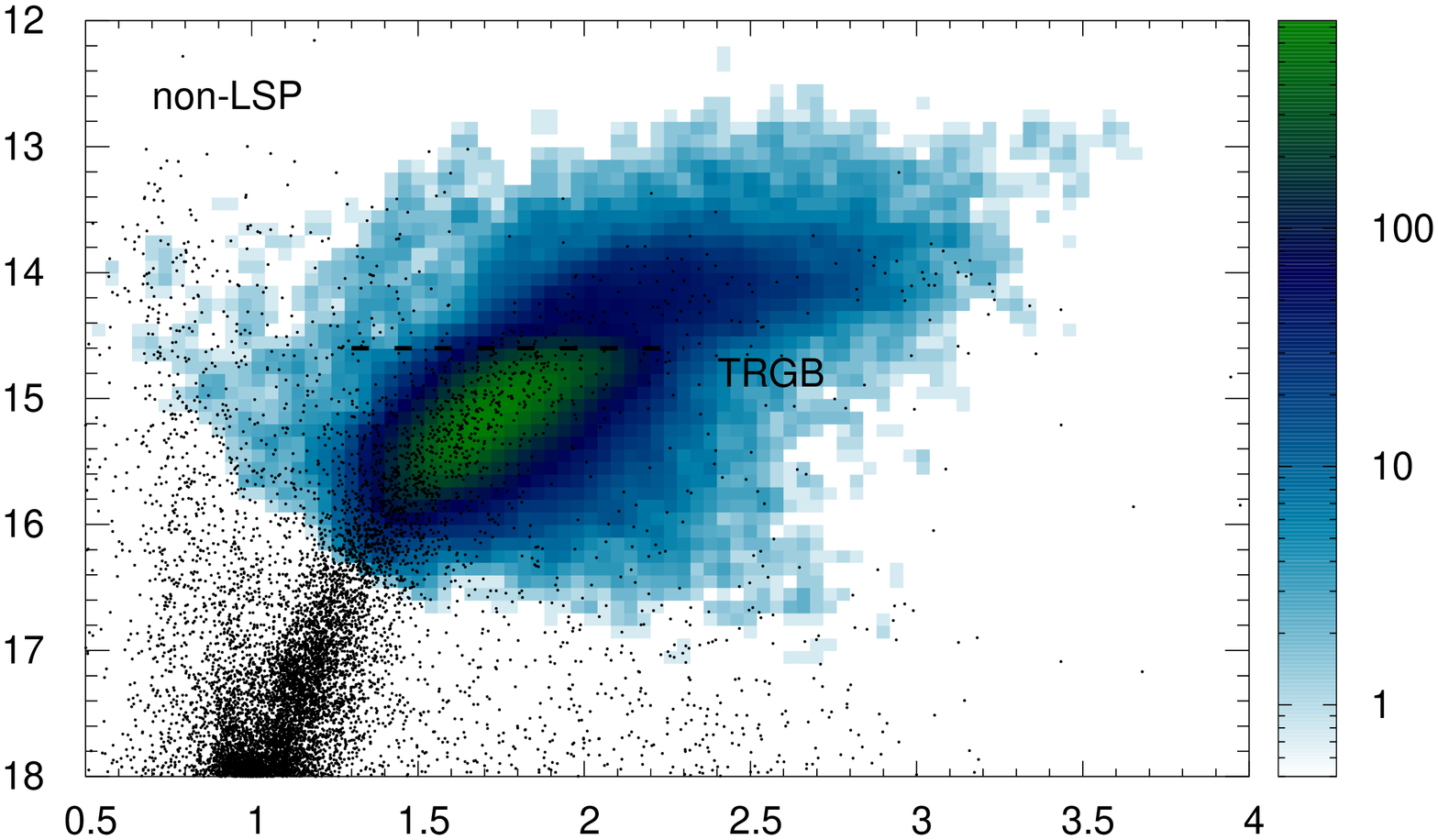} \\
	\includegraphics[bb=100 50 480 800, width=0.41\textwidth, angle=270, clip]{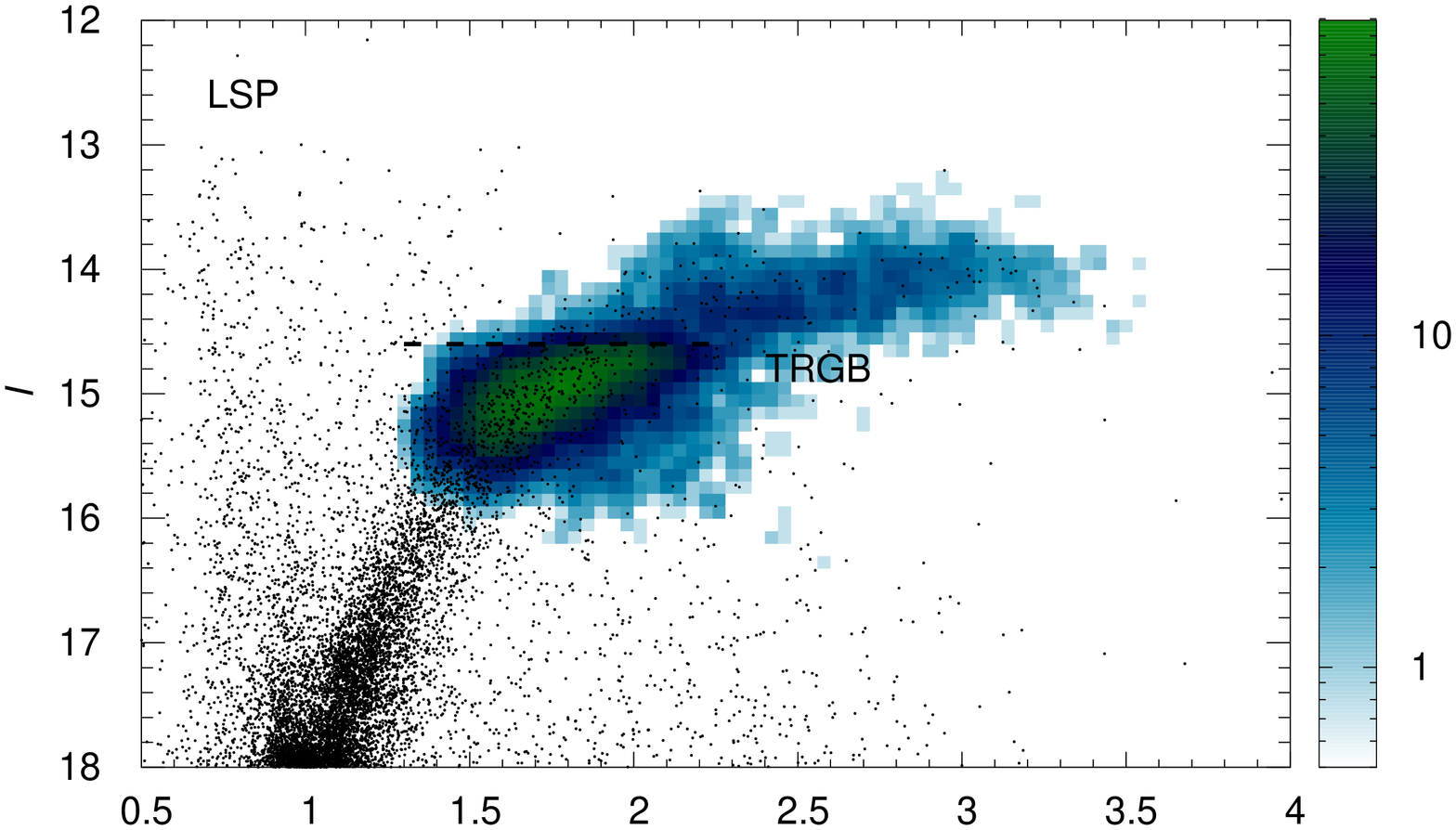} \\
	\includegraphics[bb=100 50 535 800, width=0.47\textwidth, angle=270, clip]{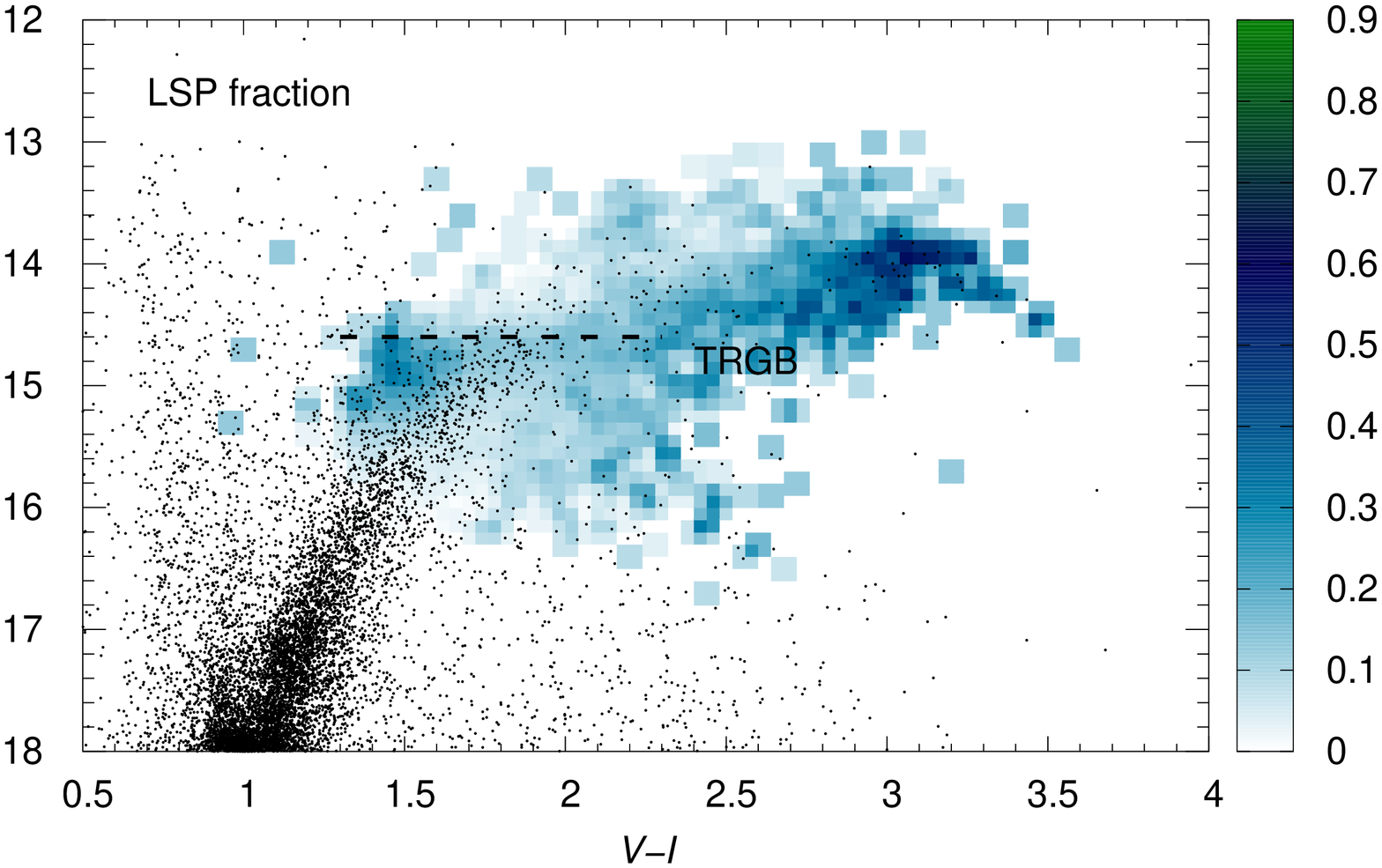}
   \caption{Density map in the CMD plane. Panels as in Fig.~\ref{fig:pl}. The upper and middle panels are plotted in log scale. The black dots are LMC field stars from the OGLE photometric map \citep{udalski2008}.}
    \label{fig:cmd}
\end{figure*}

\begin{figure*}
 \centering
	\includegraphics[bb=100 50 480 800, width=0.41\textwidth, angle=270, clip]{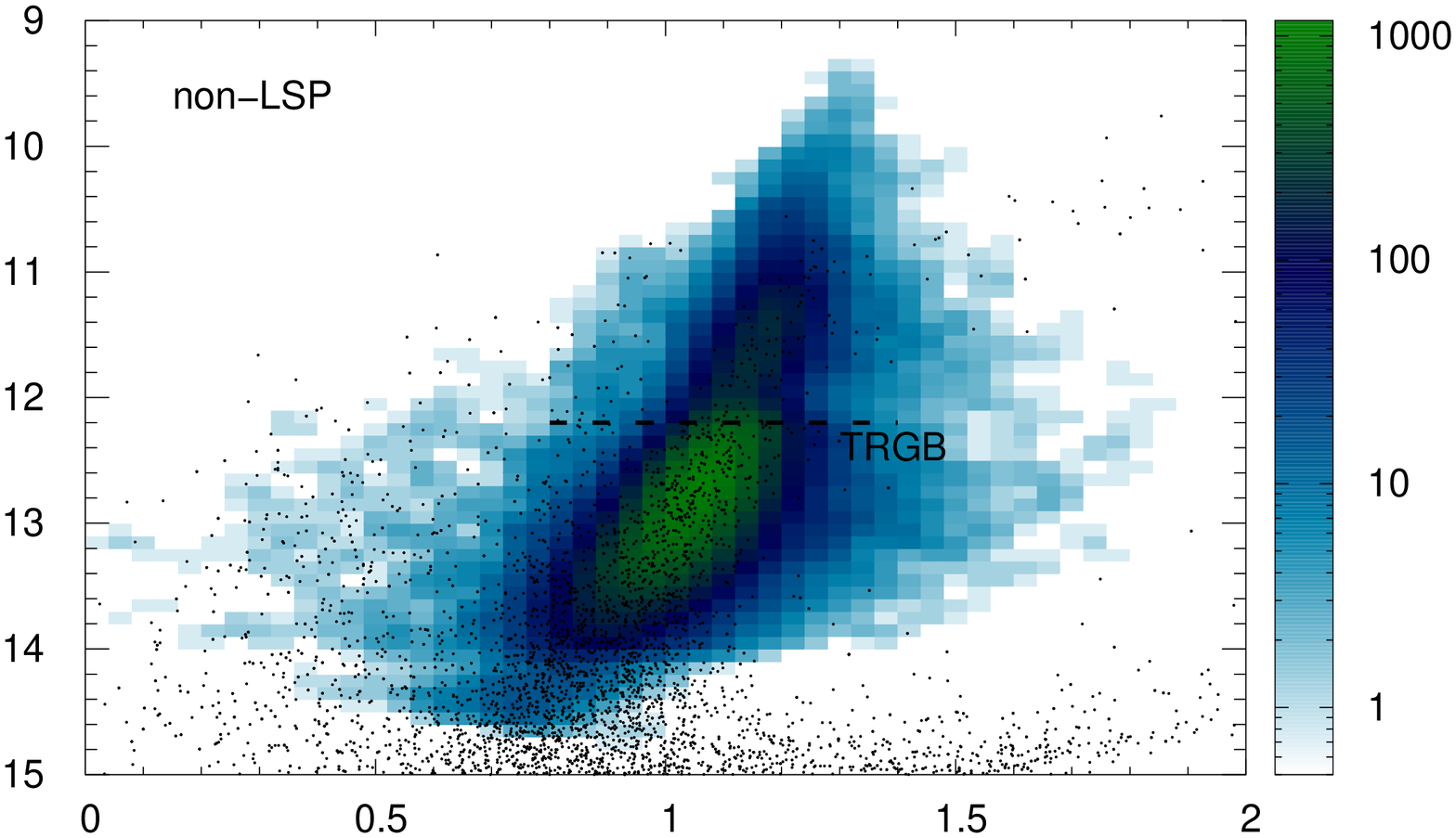} \\
	\includegraphics[bb=100 50 480 800, width=0.41\textwidth, angle=270, clip]{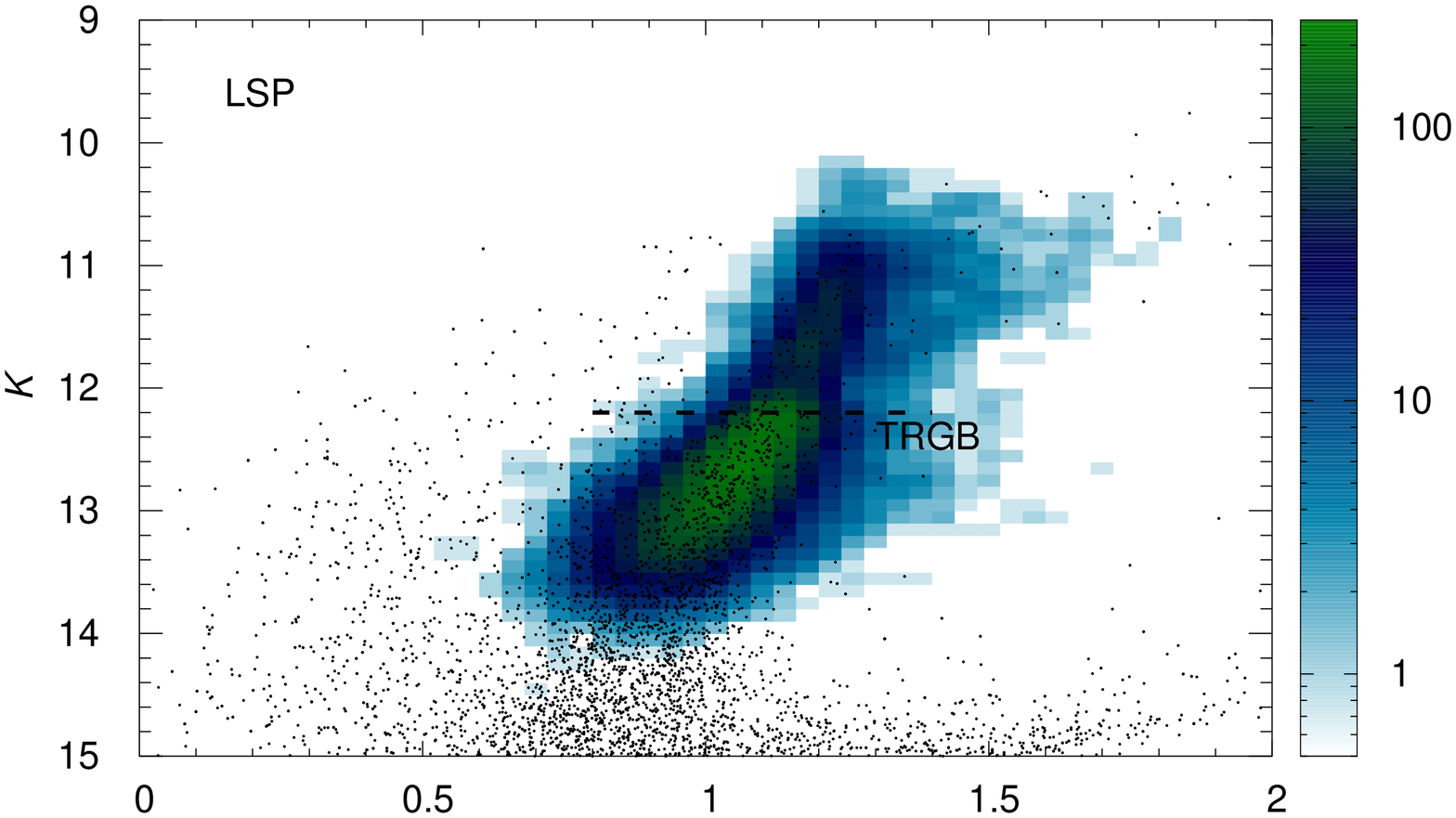} \\
	\includegraphics[bb=100 50 535 800, width=0.47\textwidth, angle=270, clip]{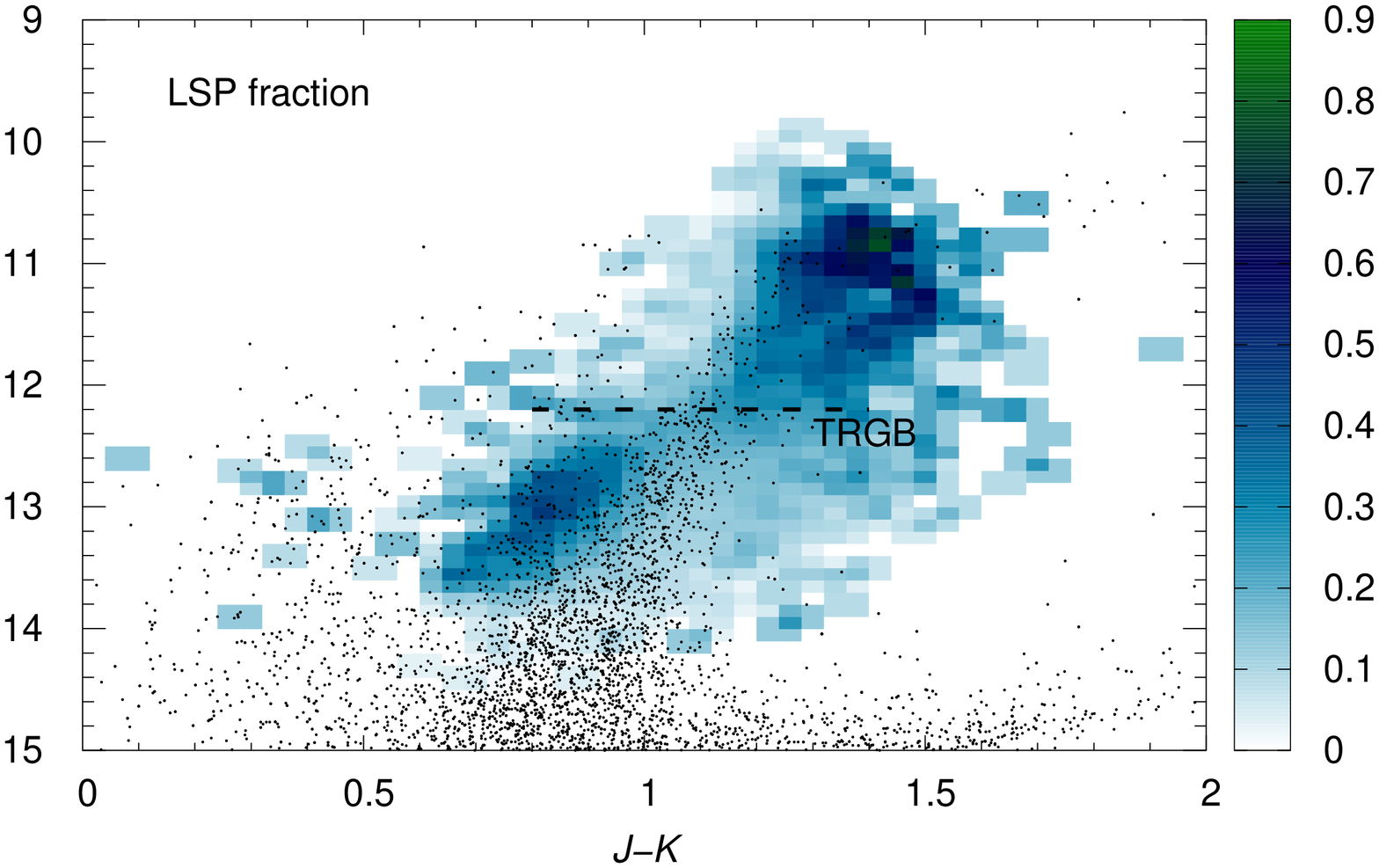}
   \caption{The same as Fig.~\ref{fig:cmd} but in 2MASS colors. }
    \label{fig:cmdk}
\end{figure*}

Fig.~\ref{fig:cmd} presents more maps, this time in the color-magnitude diagram (CMD). For both non-LSP (Fig.~\ref{fig:cmd} upper panel) and LSP (Fig.~\ref{fig:cmd} middle panel) stars, the bulk of the objects is located in the region where both RGB and AGB stars co-exist. 

The plot of the relative density of the LSPs compared to all OSARGs reveals interesting features of the LSP distribution. There seem to be two local LSP over-density regions. The first, less prominent one is located near the Tip of the Red Giant Branch (TRGB), below $I = 14.6$~mag which is adopted as the intrinsic $I$-band magnitude of the TRGB by \citet{freedman2019}. It appears to be slightly shifted to the blue side in respect to the bulk of the RGB. The second relative over-density region is clearly located at the AGB, at the maximum luminosity reached by the OSARGs. Interestingly, this maximum appears to be shifted to fainter and redder side of the AGB. This may suggest a presence of an additional, intrinsic source of reddening. The same pattern of LSP over-density regions can be seen in Fig.~\ref{fig:cmdk}, where 2MASS $J$ and $K$ bands are used. The mean intrinsic TRGB magnitude of $K = 12.2$~mag is adopted based on \citet{freedman2020}. Again, the two relative maxima can be observed, one near the TRGB and one to the right of the AGB.

\begin{figure*}
 \centering
	\includegraphics[bb=100 50 535 800, width=0.47\textwidth, angle=270, clip]{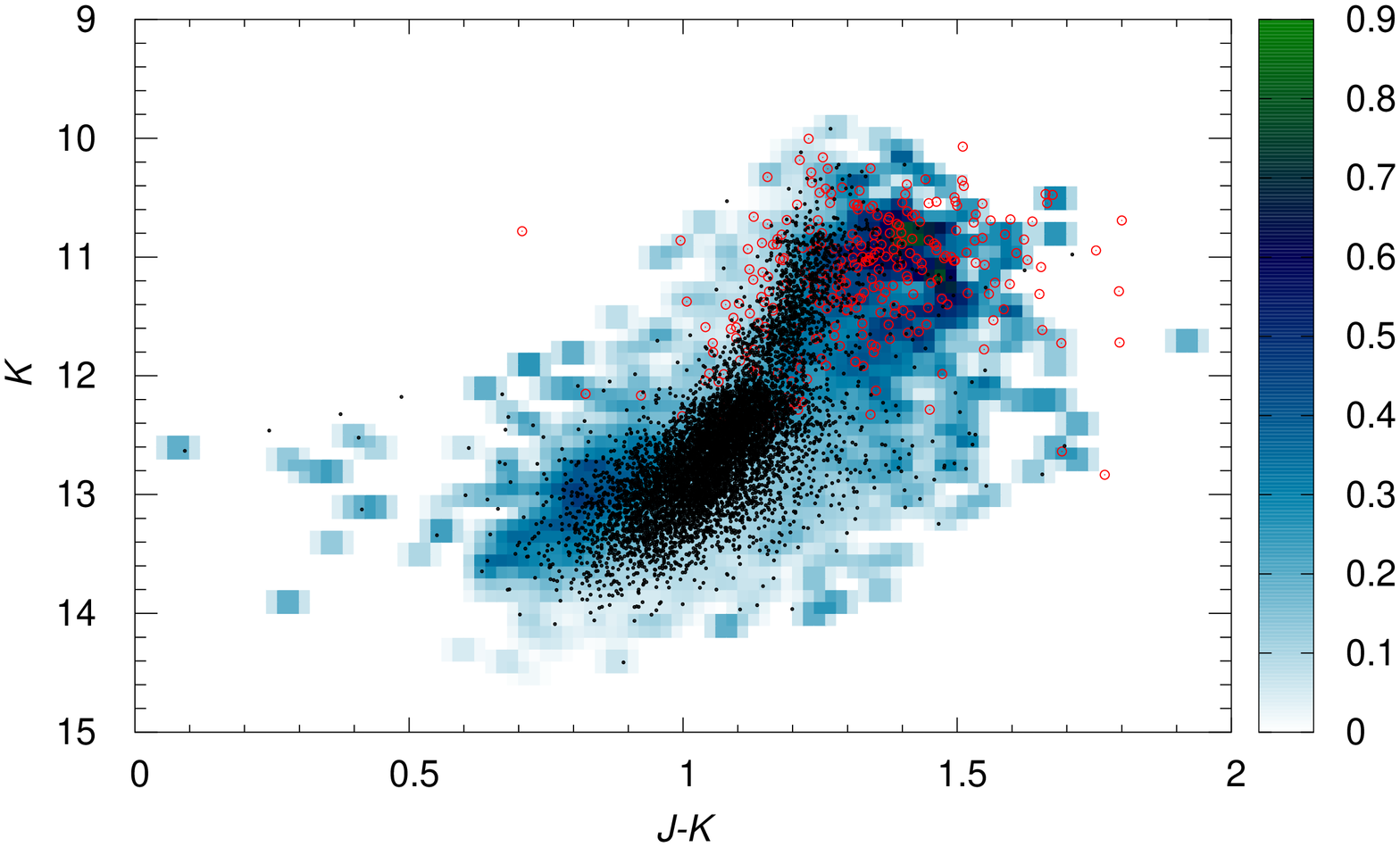}
   \caption{O- and C-rich LSP stars plotted over the LSP relative density map. O-rich stars are marked in black and C-rich - in red. The C-rich stars lie in the region where the LSP over-density appears. The O/C-rich classification comes from \citet{soszynskietal2007}}.
    \label{fig:crich}
\end{figure*}

In the lower panel of  Fig.~\ref{fig:cmdk}, the global maximum on the AGB seems to appear in the region of the CMD, where the transition from the O-rich to the C-rich stars happens. The vast majority of OSARGs are O-rich, only less then 5\% of the sample is C-rich. This is illustrated in Fig.~\ref{fig:crich} where the studied sample of the LSPs, divided into O- and C-rich, is over-plotted on the relative LSP density map.

\begin{figure*}[ht]
 \centering
	\includegraphics[bb=70 50 530 760, width=0.45\textwidth, angle=270, clip]{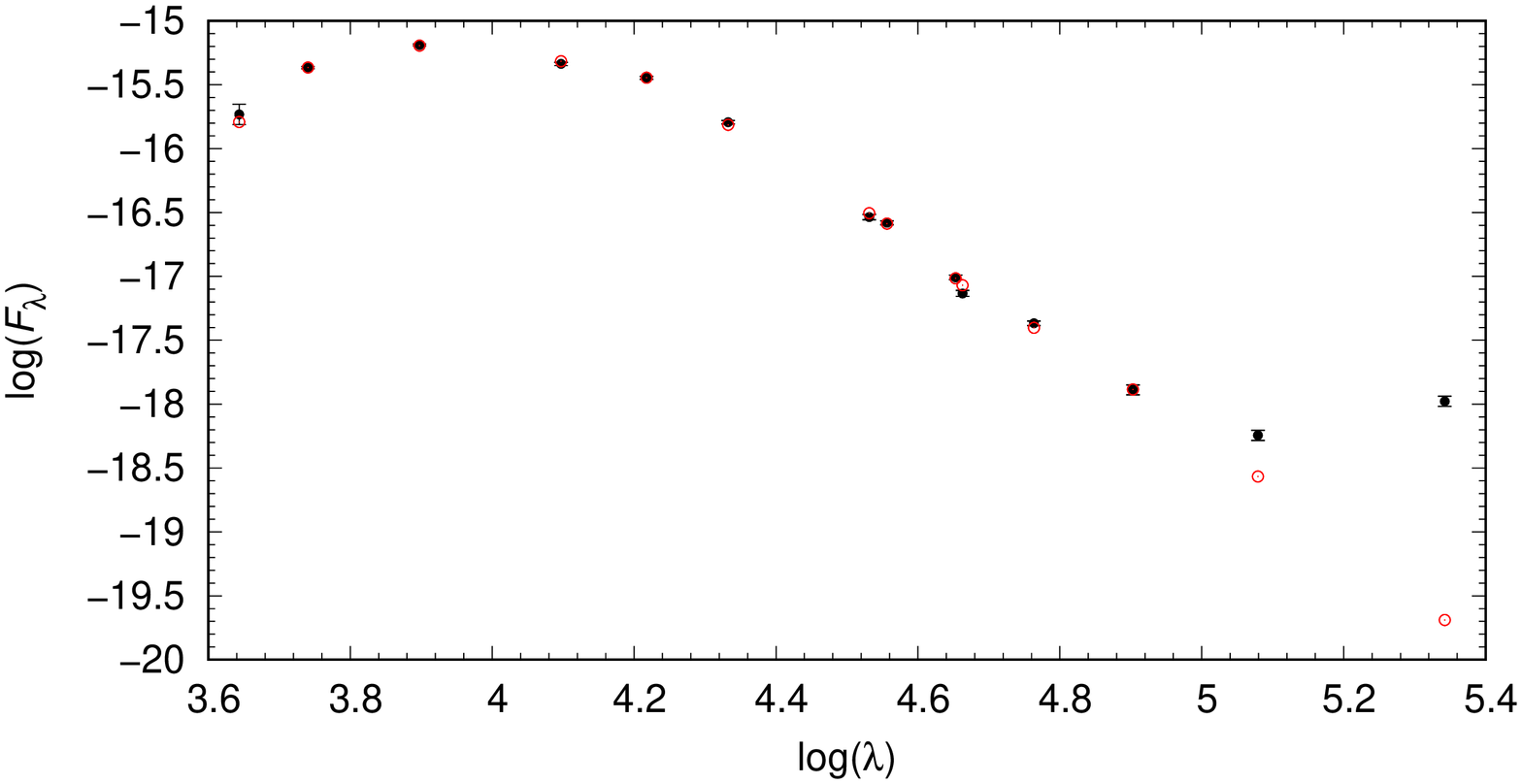} 
	\caption{An example of the SED fit performed for OGLE-LMC-LPV-66712. The observations are marked with black solid circles with error bars and the model with red empty circles. The two most red points (W3 and W4) were ignored during the fitting.}
    \label{fig:sed}
\end{figure*}

\begin{figure*}[ht]
 \centering
	\includegraphics[bb=40 50 560 760, width=0.55\textwidth, angle=270, clip]{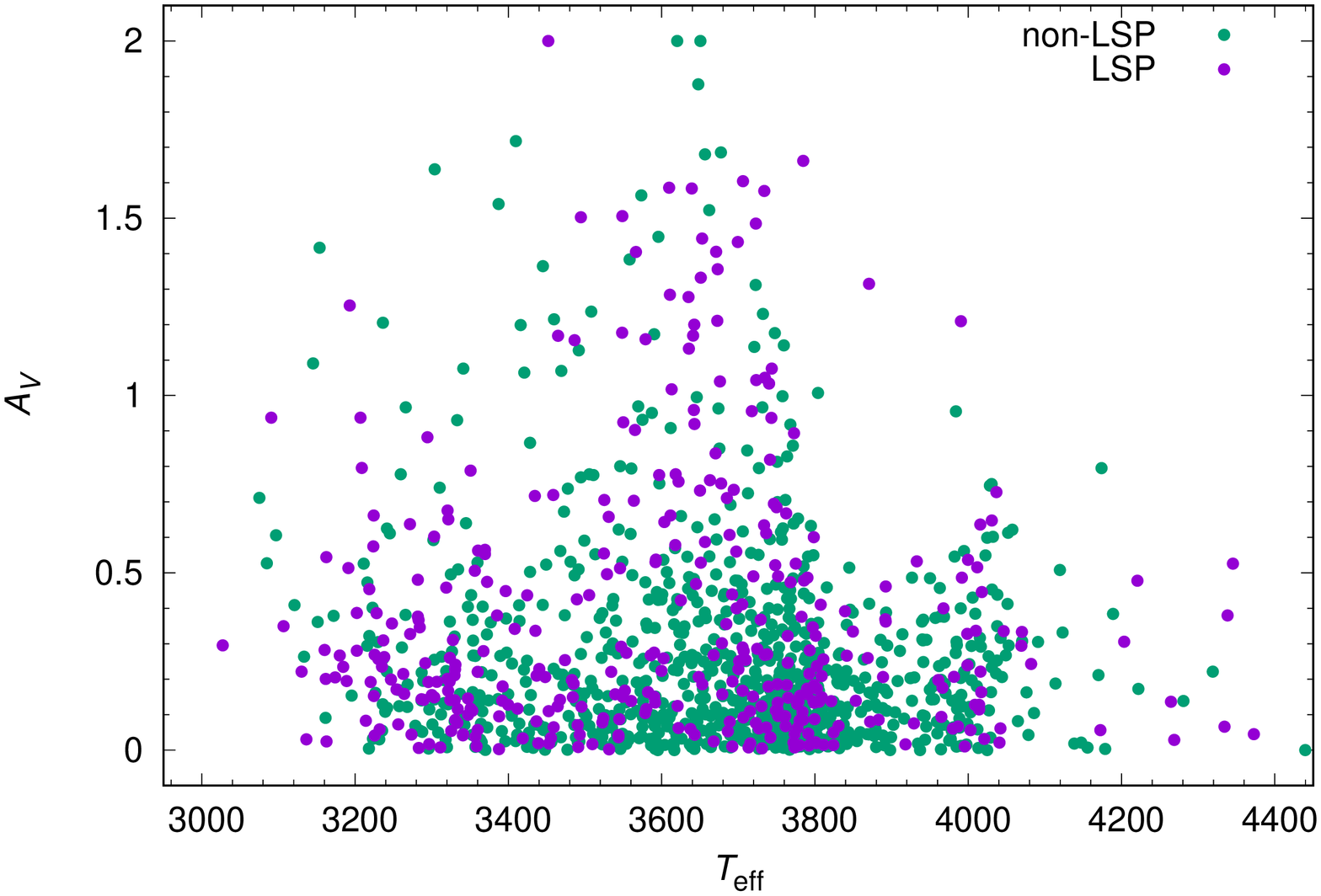} 
	\caption{Distribution of the sample in the $T_{\rm{eff}}$ vs. $A_V$ plane. No obvious correlation between $T_{\rm{eff}}$ and $A_V$ is visible.}
    \label{fig:teffav}
\end{figure*}

\citet{wood2009} have already found excess mid-infrared emission from the LSP stars in the LMC using Spitzer mid-infrared data. To further verify the hypothesis about the additional reddening of the LSP stars, I analyze the spectral energy distribution (SED) of the LSPs and non-LSPs, using the photometric data from OGLE (V and I-bands from OGLE-III and B-band from OGLE-II), 2MASS J, H, K bands \citep{skrutskie2006}, Spitzer  \citep{werner2004} and Wise \citep{wright2010} bands. The magnitudes are converted to fluxes assuming the zeropoints provided by \citet{bessell1979} for the OGLE data and the the zeropoints provided by in the surveys themselves for the remaining filters.

The observational SEDs, corrected for the extinction using \citet{cardelli1989} reddening law, are then fitted with theoretical models. The NextGen Model grid \cite{allard2012} is used for that purpose. The total extinction value $A_V$ is treated as a free parameter of the fit, assuming the mean $R_V = 3.1$. The models assume $\log g = 0.5$, [Fe/H] $= -1$ and a grid of effective temperature ($T_{\rm{eff}}$) from 3000 to 4500~K, with a 100~K step. Due to the strong correlation between $A_V$ and $T_{\rm{eff}}$, the 100~K step provided by NextGen is insufficient to derive $A_V$ accurately enough. Therefore, I need to implement the interpolation of the model to be able to obtain the values of the flux in all the passbands used for the fitting, for an arbitrary temperature. This is done with a cubic interpolation in $F_{\lambda}$ - $T_{\rm{eff}}$ space, for each passband. 

\begin{figure*}
 \centering
	\includegraphics[bb=130 60 460 760, width=0.35\textwidth, angle=270, clip]{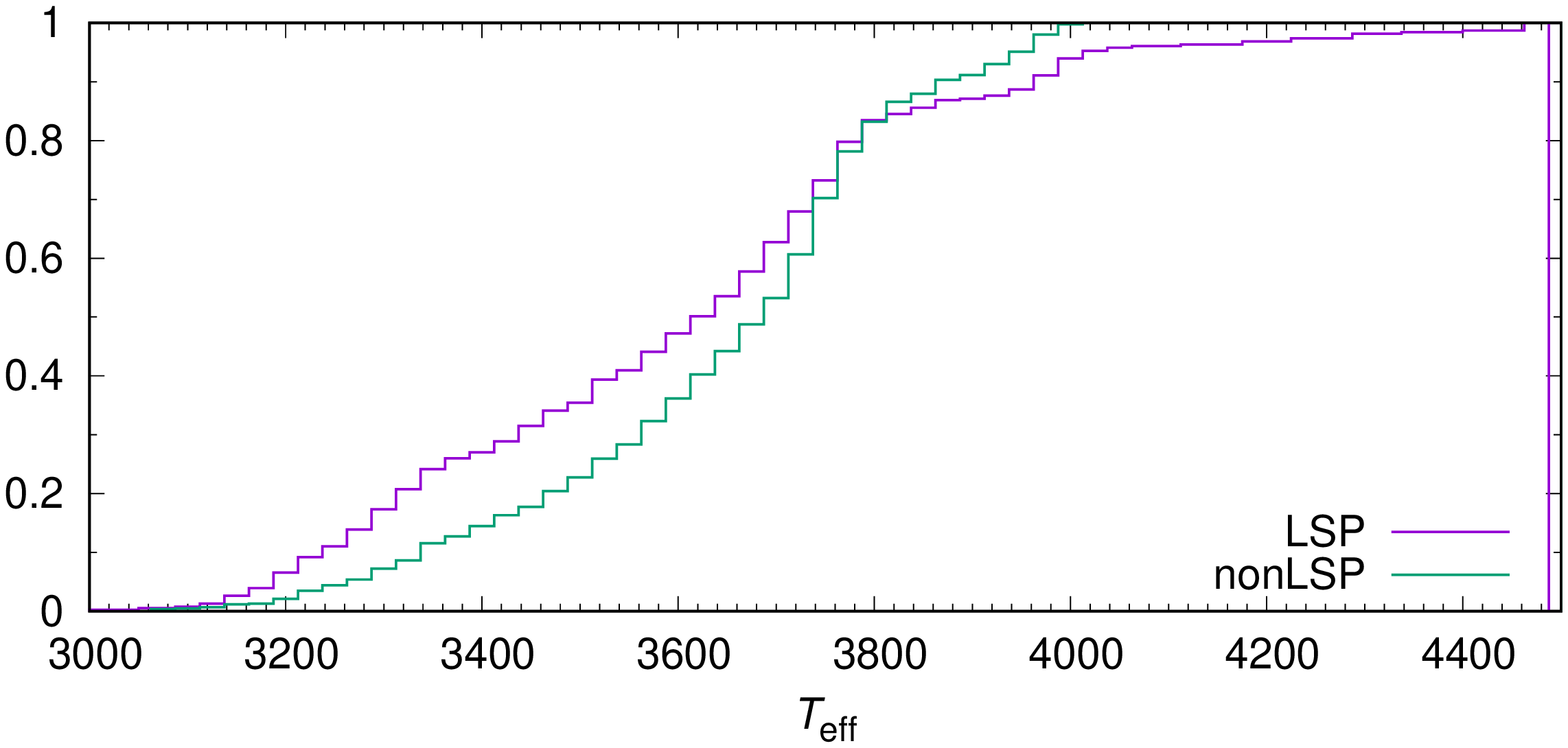} 
	\caption{Cumulative distribution of $T_{\rm{eff}}$ derived with the SED fitting for LSP and non-LSP stars.}
    \label{fig:histteff}
\end{figure*}

\begin{figure*}[ht]
 \centering
	\includegraphics[bb=130 60 460 760, width=0.35\textwidth, angle=270, clip]{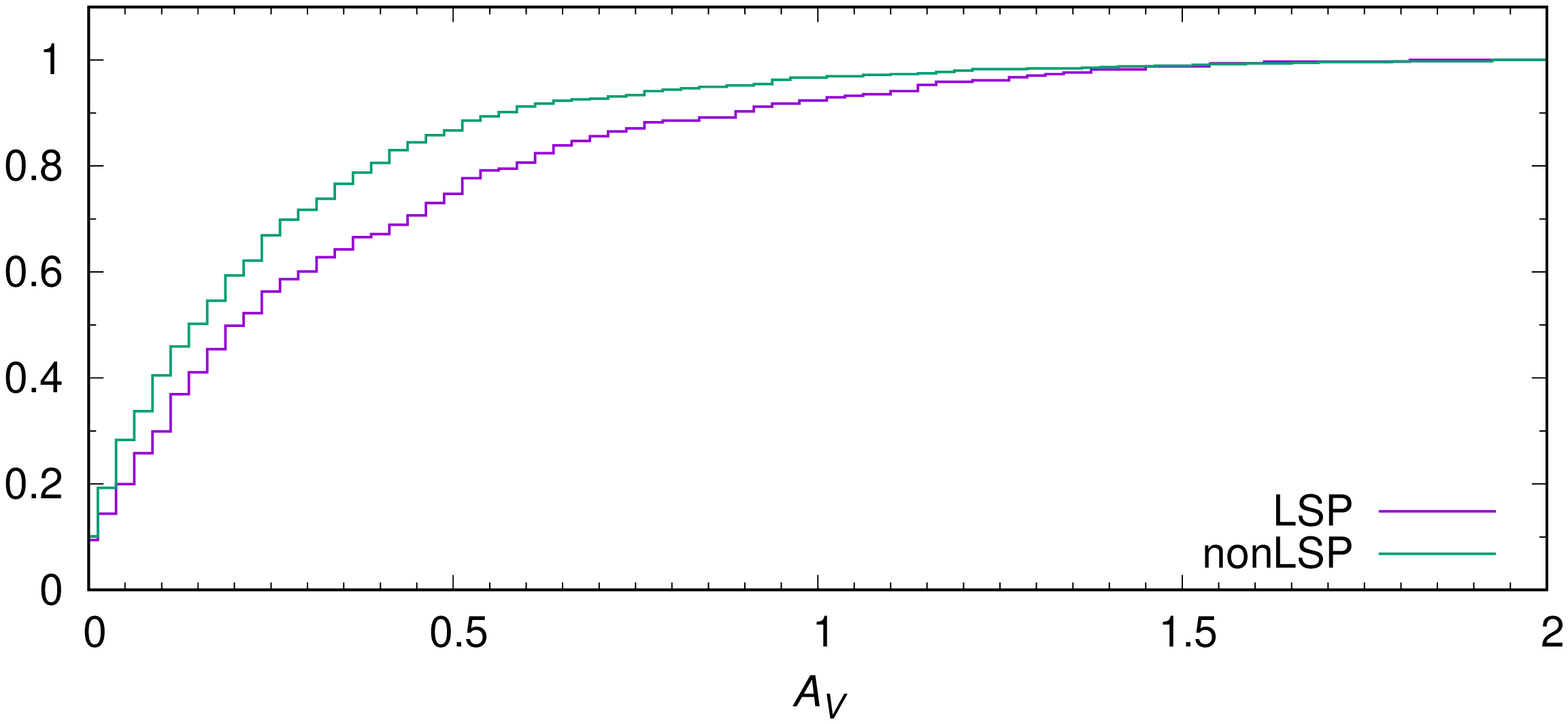} 
	\caption{Cumulative distribution of $A_V$ derived with the SED fitting for LSP and non-LSP stars.}
    \label{fig:histav}
\end{figure*}

Out of the whole sample, I extract a subset of 1310 LSP and 5552 non-LSP stars, for which the measurements in all pass bands are available. For these stars I fit the $T_{\rm eff}$ and $A_V$ simultaneously. An example of the fit is presented in Fig.~\ref{fig:sed}. Out of the entire sample only 381 LSP and 858 non-LSP stars got fitted with $A_V > 0$. This is a common problem with SED fitting that has been reported for example by \citet{zaritsky2004}. It is likely to be related to various factors, including degeneracy between $A_V$ and $T_{\rm eff}$ and difficulty with a precise determination of the uncertainties of the photometric measurements, as well as the fact that the single-epoch magnitudes are used for objects that are strongly variable in time. Being aware of the limitations of this approach, I still consider it useful for the studied problem, even though the results should be interpreted with caution.  
 
To compare the distribution of the reddening, I reject all the objects with nonphysical results of the fit and take into account only the systems with positive $A_V$. As a first step, I check for a possible correlation between $T_{\rm{eff}}$ and $A_V$. For that purpose, I construct Fig.~\ref{fig:teffav}, which does not show any strong correlation. The density of the LSP stars seems to be higher at the low-temperature end of the diagram, which is confirmed in Fig.~\ref{fig:histteff}, showing the cumulative distribution in $T_{\rm{eff}}$. This is consistent with the observational data (Fig.~\ref{fig:cmd} and \ref{fig:cmdk}) where it can be seen that the fraction of the LSPs increases while moving to the redder, and therefore cooler, part of the CMD. 

The cumulative distributions for both LSP and non-LSP stars in $A_V$ is shown in Fig.~\ref{fig:histav}. It appears that the LSP stars have on average higher $A_V$ value. This is consistent with the hypothesis about an additional, intrinsic source of reddening present in those objects. However, taking into consideration the large uncertainty of the SED fit related to the aforementioned factors, this result can be interpreted only as a hint, not as a self-standing proof.

\section{Discussion}

The LSP phenomenon may be connected to the transition between different PL sequences. The over-density of the LSP stars between the sequences $a_2$ and $a_1$ has been already noticed by \citet{trabucchi2017}. This analysis reveals that the appearance of the LSP phenomenon may not be restricted to this particular transition, but can be also observed for stars lying between sequences $a_3$ and $a_2$ and sequences $C'$ and $C$. The number of object between the last two sequences is very low, however the relative density of the LSP stars is very high. The various PL sequences are generally attributed to different pulsation modes changing from the higher overtones at the short period end to the fundamental mode at the long period end \citep{trabucchi2019}, however, the attribution of a particular sequence to a given overtone varies between authors. 

While evolving on the RGB and later on the AGB, a red giant increases both its luminosity and period \citep{wood2015}, moving from the bottom left to the top right on the PL diagram (Fig.~\ref{fig:pl}). During its evolution the giant can cross various PL sequences, switching its dominating pulsation mode from the higher to lower overtone, to finally reach the fundamental mode. 
This means, the giant can also cross more than one region, where the LSP becomes prominent in the PL plane. 
This suggest that the LSP may not be a single time episode but a recurring phenomenon, which can appear and disappear multiple time for a given star. 

The fact that the LSP emerges mostly in between the PL sequences may suggest, that the mechanism that triggers it is related to the switching of the dominating pulsation mode. Since the pulsational periods are becoming longer in the course of evolution, the stars undergoing the dominant mode transition should be on the long period side of the sequence representing the
old dominant mode, or on the short period side of the sequence representing the new one. The combined amplitude of decaying higher- and emerging lower-overtone modes can be a key factor in starting the LSP. This could explain the offset of the LSP stars from the PL sequences.

However, it should be noted that the sequences $a_2$ and $a_1/C'$ are both interpreted as the same, first overtone sequences by \citet{trabucchi2017}. In that case, the main LSP over-density region lying between these sequences could not be explained by a dominant pulsation mode transition. 

While the offset of the LSP stars from the PL sequences is quite clear between $a_3$ and $a_2$ as well as $C'$ and $C$, it is not obvious whether such offset, in reference to the non-LSP OSARGs, can be observed between $a_2$ and $a_1$. \citet{trabucchi2017} shows that the stars with with dominant first overtone pulsation often have a second overtone lying on the long period side of the sequence $b_3$. The reason for the offset is the mass difference between the typical stars on the different sequences \citep{wood2015}. As OSARG stars are multi-modal pulsators, it is possible that stars forming the local maxima of the LSP to the right from the sequence $a_3/b_3$ and $C'$ are stars from the main $a_2/a_1$ over-density region that where simply detected with a different overtone.

On the other hand, even with the assumption that the LSP appears only in stars that pulsate predominately in the first overtone, the presence of a LSP can still be possibly tied to the dominant mode transition. The fact that a significant fraction of the LSP stars, which should be detected with the first overtone, were actually detected with the second overtone as the strongest pulsational mode suggests that while the first overtone is already excited, the second overtone may not have full decayed yet, which may hint to a recent or still ongoing mode transition. However, in this scenario there would be just a single transition, not multiple ones. 

A significant fraction of LSP stars lie above the TRGB, making it clear that the LSP population contains AGB stars. The presence of RGB stars in the LSP sample is less obvious, due to overlapping of the RGB and AGB regions of the CMD. However, the noticeably drop in the LSP density above the TRGB as well as the local maximum of the relative LSP density, strongly suggests that the RGB stars are also present in the LSP sample, and the population of the LSP giants below the TRGB is a mixture of RGB and AGB stars. The presence of more than one maximum in the LSP relative density: the local maximum around the TRGB and the global maximum on the AGB, also seems to be consistent with the re-occurring LSP scenario.

The global maximum of the relative LSP density corresponds to the upper part of the AGB, in this region there are areas where more than half of the OSARGs show LSP. This maximum appears to be shifted to the red in color and fainter in magnitude in respect to the AGB. This hints that the LSP stars are more reddened than the rest of the AGB population. The fact that the LSP population is more reddened than rest of the giants, suggest that the cause of the reddening must be intrinsic to the stars themselves. \citet{mcdonald2019} suggested a connection between a high mass loss rate, and therefore large dust production, and the emergence of LSP. The higher reddening of the LSP stars, which can be caused by presence of a dust cloud around the giant, seems to be in agreement with that. 

The mass loss rate is known to increase during the evolution on both RGB and AGB \citep{schroder2005, ramstedt2014}, similarly to the LSP fraction which also rises while moving along the AGB on the CMD. Recent study by \citet{yasuda2019} investigating the different types of wind in red giants shows that a stable wind is observed only at the end of the RGB evolution near the TRGB, as well as at the end of TP-AGB phase at the end of AGB evolution. This two evolutionary phases coincide with the local and global maximum of the relative density of the LSP stars on the CMD diagram. This is another strong indication that the LSP mechanism is tied to the presence of the wind and mass loss.

While the connection between the appearance of LSP and stellar wind, mass loss and dust production seems to be clear, the mechanism behind LSP still remains unknown. Especially, it is not clear whether LSP is the cause or the result of the strong mass loss. It might be, as suggested by \citet{mcdonald2019}, that the strong mass loss is triggered by the presence of an additional pulsation mechanism attributed to LSP. On the other hand, assuming that LSP is related to the binarity and the presence of a low-mass companion that drags behind it a dust cloud, which periodically obscures the giant, the high mass loss would play a key role in triggering such mechanism by providing the dust necessary to form a dust cloud around the low-mass companion.

\section{Conclusions}

In this work, I present a hypothesis that the LSP phenomenon can be related to the transition between the pulsation PL sequences. It does not have to be tied to a single region of the PL diagram, but appears in three areas - between sequences $a_3$ and $a_2$, $a_2$ and $a_1$, as well as $C'$ and $C$. As particular PL sequences correspond to different pulsation mode, the appearance of LSP may connected with the change of the dominant pulsation mode. However, it might be that the two local over-density regions are simply formed by the same stars as the ones from the main over-density between $a_2$ and $a_1$, when such stars are detected with a different pulsation mode. The main over-density region was already reported by \citet{trabucchi2017} and was not interpreted as a mode transition region, since both $a_1$ and $a_2$ sequences were attributed to the first overtone pulsation.

The existence of multiple regions where LSP is prominent suggests that LSP may not be single-time episode, but a recurring phenomenon appearing and disappearing during the red giant evolution. The existence of the two maxima of the LSP relative density appears to be consistent with this explanation.

In the AGB-only region above the TRGB, I also notice that the main over-density region of the LSP on the CMD appears to be shifted towards the fainter and redder direction, which is likely a result of a stronger reddening of these objects. If the reddening is specific for LSP stars only, it must be related to some phenomenon intrinsic to these objects - most likely absorption by the dust in the system. The results of the SED fitting seem to confirm the reddening excess in the LSP stars,  however, due to the limitation of the method this result should be interpreted with caution.

The presence of the LSP over-densities in the CMD regions corresponding to the evolutionary stages on RGB and AGB where stable winds and strong mass loss are observed is another indication of relation between LSP and high mass loss. Whether the mass loss is a trigger or a cause of LSP, remains unknown.

\begin{acknowledgements}
I would like to thank the Referee for insightful comments.
I also thank Ond{\v{r}}ej Pejcha and Igor Soszy{\'n}ski for fruitful discussion. 
This work has been supported by the INTER-EXCELLENCE grant LTAUSA18093 from the Czech Ministry of Education, Youth, and the SONATINA grant 2020/36/C/ST9/00103 from the Polish National Science Center. This research made use of Astropy, a community-developed Python package for Astronomy \citep{astropy}.
\end{acknowledgements}


\end{document}